\documentclass[final, 1p, authoryear]{elsarticle}
\usepackage{pdfwidgets}
\usepackage{enum}
 \usepackage{amsmath}
 \usepackage{here}
\usepackage[T1]{fontenc}
\usepackage{amsbsy}
\usepackage{latexsym}
\usepackage{arydshln,booktabs,bm}
\usepackage{dcolumn,warpcol}  

\journal{Journal of Multivariate Analysis}

\newcommand{\logit}{\mbox{\rm logit}\,}

\newcommand{\fracshalf}{\mbox{$\frac{1}{2}$}}

\newcommand{\fracsfourth}{\mbox{$\frac{1}{4}$}}

\newcommand{\txt}{\textstyle}
\newcommand{\T}{^{\mathrm{T}}}

\newcommand\cov{\mathrm{cov}}
\newcommand\var{\mathrm{var}}

\newcommand{\E}{{\it E}}

\newcommand{\bcal}{\mathcal{B}}
\newcommand{\ecal}{\ensuremath{\mathcal{E}}}

\newcommand{\tcal}{\ensuremath{\mathcal{T}}}
\newcommand{\ical}{\ensuremath{\mathcal{I}}}

\newcommand{\ci}{\mbox{\protect{ $ \perp \hspace{-2.3ex}
\perp$ }}}

\newcommand{\n}[0]{\hspace*{.35em}}

\newcommand{\nn}[0]{\hspace*{.7em}}

\newcommand{\fourl}[0]{\hspace*{1.4em}}

\begin{document}
\vspace{-1cm}
\noindent Journal of Multivariate Analysis 130 (2014) 252--260; \n also  on ArXiv: 1311.5655\\[3mm]

\noindent{\Large \bf Binary distributions of concentric rings}\nn \\[6mm]
{\large \bf  Nanny Wermuth\footnote[1]{Corresponding author: wermuth@chalmers.se} \\[2mm]}
{\it Departm.\  of Math. Sciences, Chalmers University of Technology, Sweden, and
Departm.\ of Medical Psychology and Medical Sociology, Johannes Gutenberg-University, Germany
\\[4mm]}
{\large \bf Giovanni M. Marchetti\\[2mm]}{\it Dipartm. $\! \!$di Statist.$\!$,$\!$ Informatica, Applicazioni,  ``G. Parenti'', 
University of Florence, Italy\\[4mm]}
{\large \bf Piotr Zwiernik   \\[2mm]}
{\it Department of Statistics, University of California, Berkeley, USA}  \\[1mm]

{\em \noindent{\bf Abstract:} We introduce families of jointly  symmetric, binary distributions that are generated over directed star graphs whose nodes represent variables and whose edges indicate positive dependences.  The families are  parametrized in terms of a single parameter.  It is an outstanding feature of these distributions that joint probabilities relate to evenly spaced concentric rings.  Kronecker product characterizations make them computationally attractive for a large number of variables. We study the behaviour of different measures of dependence and derive maximum likelihood estimates when all nodes are observed and when the inner node is hidden.} \\

\noindent\textit{Key words}:  {\rm  
Conditional independence; Graphical Markov models; Jointly symmetric distributions; 
Labeled trees; Latent class models;  Phylogenetic trees; Star graphs; Symmetric  variables.}

\section{Introduction}

We define and study  a family of distribution for $p=1, 2, \ldots $ binary random variables, denoted by $A_1, \ldots , A_Q, L$. Each variable has equally probable levels, so that the variables are symmetric. There are $Q$ response variables  $A_1, \ldots , A_Q,$ to a single
common explanatory variable $L$, named the signal and having the levels  strong or weak. The possible responses are to succeed or to miss.
We use as  a  convention that success  for $A_q$ is coded 1 and that a strong  signal  of $L$  is  also coded 1.  For the low level, we use either $-1$ or   $0$. Of special interest are situations in which the signal cannot be directly observed, it is instead hidden or latent, but the aim is to understand 
and estimate the joint structure including $L$.  In that case, we have $t=1, \dots, 2^Q$ level combinations.

 We  let $K_t= a_1 + \cdots + a_Q$ denote  the number of ones in any given sequence of response-level combinations, $(a_1,\dots, a_Q)$, and
 define a  normalizing constant, $c_Q=2(1+\alpha)^Q$  for  $1\leq \alpha < \infty$, to write with $\{0, 1 \}$ coding, also known as baseline coding, 
for  the joint $p$-dimensional distribution
 \begin{equation} \label{jointprobs}
 \pi(a_1,\dots ,a_Q, l) \, c_Q  \quad=\quad
\begin{cases}
\alpha^{K_t} & \rm{ \n for \n} l=1,\\
\alpha^{(Q-K_t)} &  \rm{\n  for \n} l=0 \, .\\
 \end{cases} 
\end{equation}

   For the $\{-1, 1 \}$ coding of the levels, known  also as effect coding, the symmetry of each of the  binary variables implies zero mean and unit variance.  For $L$, we write
$$ \,pr
(L=1)=\,pr
(L=-1)=\fracshalf,   \nn \E(L)=0, \nn \E(L^2)=1.$$
For any such binary variable pair $(A,L)$, the correlation coefficient $\rho$, which is 
$$ \rho\,\,=\,\,\cov(A,L)\,\,=\,\,\E(A\, L),$$
ranges in   $0 \leq \rho<1$ and 
 \begin{equation}\alpha=(1+\rho)/(1-\rho), \nn \nn  \rho=(\alpha-1)/(\alpha+1)\, . \label{rhoalpha}\end{equation}
 The correlation $\rho$ is also the regression coefficient in a projection of $A$ on $L$. Furthermore,  independence of $A$ from $L$, denoted by $A \ci L$, relates to $\alpha$ and $\rho$ via 
 $$ A \ci   L \n \iff\n (\alpha=1) \n \iff \n  (\rho=0).$$
 
   This  last case would give a degenerate model  in equation \eqref{jointprobs}, hence it is excluded for some purposes.  The following Table~\ref{tab:range} shows how 
two types of sequences of ratios for $\rho$ generate
all possible even and odd positive integers   for $\alpha$ and hence proper counts in equation \eqref{jointprobs}.

\begin{table}[H]
\centering
\caption{An integer valued $\alpha$ for symmetric binary variables in concentric-ring models}\label{tab:range} \vspace{2mm}
\small
\begin{tabular}{lllllllllllllllllll}
\toprule\\[-3mm]
$\alpha$ \nn & 1\nn & 3& 5& 7& 9& 11& 13 & 15&$\dots$\\
$\rho$ &0& 1/2& 2/3& 3/4& 4/5& 5/6& 6/7& 7/8& $\ldots$\\[1mm]
\midrule\\[-3mm]
$\alpha$ & 2 & 4& 6& 8& 10& 12& 14 & 16&$\dots$\\
$\rho$ &1/3& 3/5& 5/7& 7/9& 9/11& 11/13&13/15& 15/17& $\ldots$\\[1mm]
\bottomrule
\end{tabular}
\end{table}

As will be shown, a model   with  density given by equation \eqref{jointprobs} has several  attractive features that were not previously identified even though it is a special case of a number of models that have been intensively studied. For instance, it is a distribution generated over a labeled tree (Castelo and Siebes, 2003), hence  a  lattice-conditional-independence model  (Perlman and Wu, 1999) and  a directed-acyclic-graph model (Wermuth and Lauritzen, 1983; Pearl, 1988) or  a Markov field for binary variables (Darroch, Lauritzen and Speed, 1980), an Ising model of ferromagnetism, a  binary quadratic exponential distribution (Besag, 1974; Cox and Wermuth, 1994) and  a  triangular system of symmetric binary variables (Wermuth, Marchetti and Cox, 2009). 

With  $L$  in equation \eqref{jointprobs} unobserved, the resulting model may be regarded as a simplest case for constructing phylogentic trees;
see Zwiernik and Smith (2011), Allman et al. (2014) and the previous extensive literature in this area.  Or, it can be viewed as  a special latent-class model (Lazarsfeld, 1950; Linzer and Lewis, 2011), the one with the closest analogy 
to a  Gaussian  factor analysis model having a single factor.  

 A {\bf
 star graph} is a directed-acyclic graph with  one inner node, $L$, from which $Q$ arrows start and point to the uncoupled, outer nodes, $1, \dots, Q$.  For $p=6$, the left of Figure 1 shows such a star graph,  having equal  regression coefficients  $\rho$ when regressing each $A_q$ on $L$, for $q=1, \dots, Q$.

\begin{figure}[H]
\centering
 \nn \nn   \includegraphics[scale=.36]{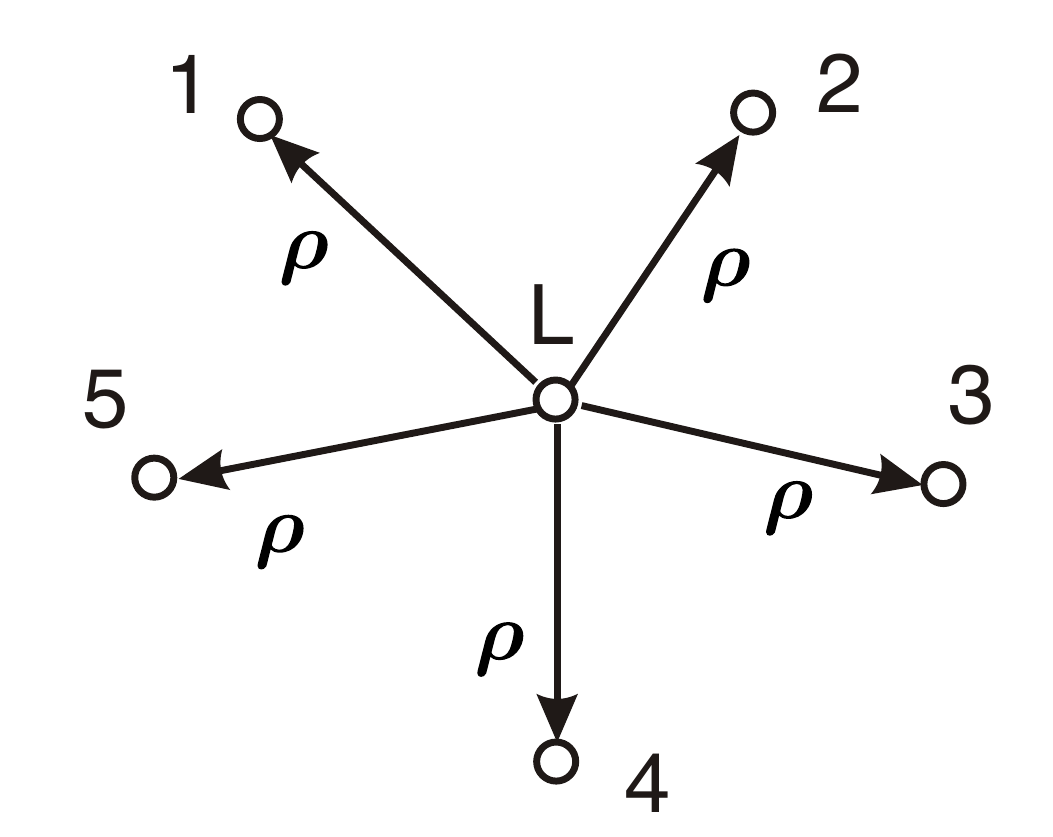}  \fourl \fourl  \fourl \fourl  \includegraphics[scale=.22]{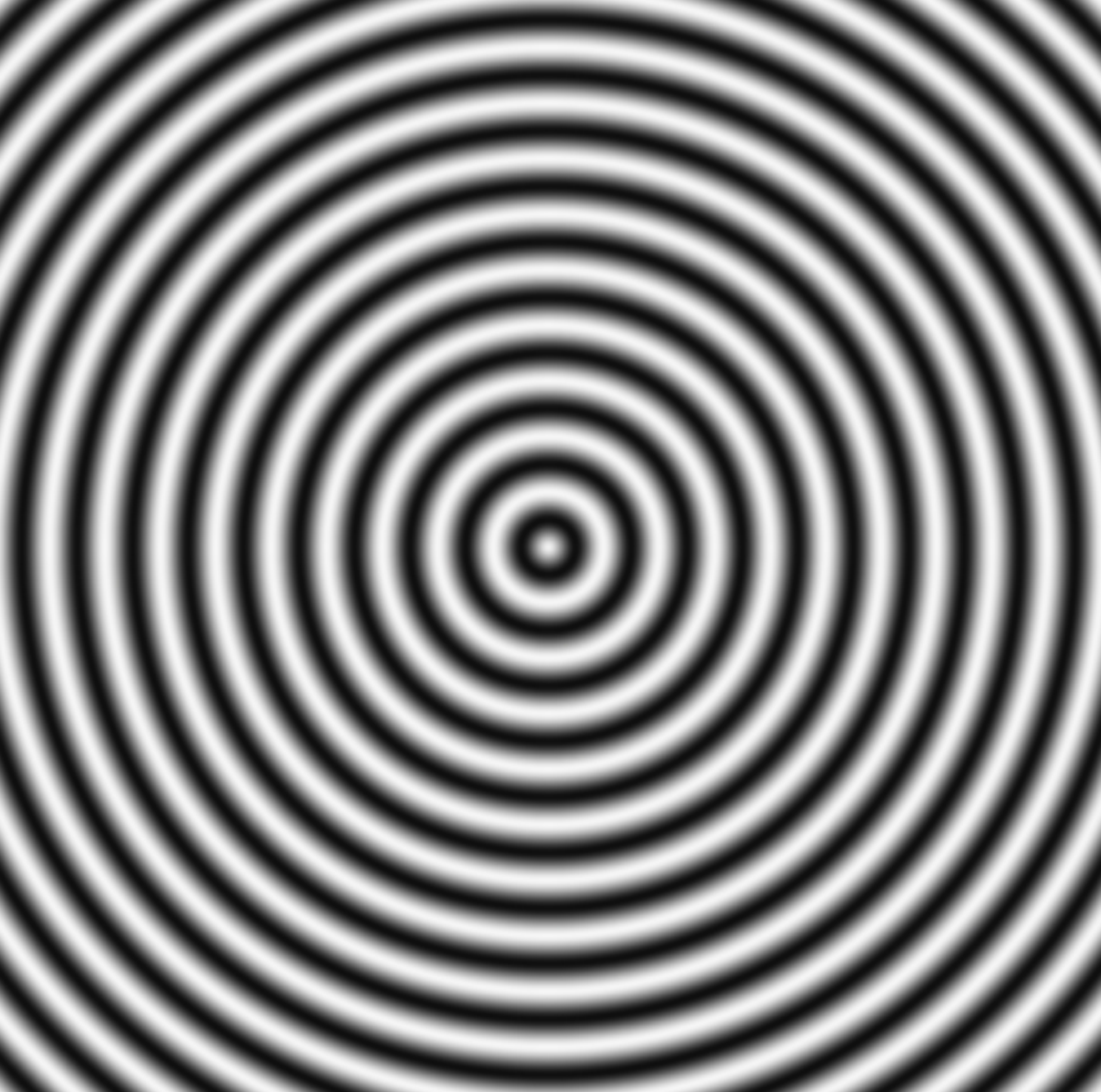}
\caption{Star graph with equal dependences of five leaves on one common root (left) and a graph of evenly-spaced concentric rings (right).}  \label{fig:1}
\end{figure}

For Gaussian and for binary distributions generated over star graphs as those in
Figure 1, the correlation matrices of the $p$ variables are of identical form;   see Wermuth and Marchetti (2014). For $p=5$, such correlation matrices are 
in Table~\ref{tab:cormat}, with `$.$' indicating a symmetric entry.
\begin{table}[H]
\caption{Correlation matrix for $p=5$; left: to equation \eqref{jointprobs}, right: to a binary latent class model} \label{tab:cormat}
 \vspace{2mm}
\centering
 $
\begin{pmatrix} \n 1\n & \n \rho^2 \n & \n  \rho^2 \n &\n   \rho^2 \n&  \rho\\
                         .& 1   &  \rho^2 &  \rho^2 &  \rho\\
                         .& .& 1 &  \rho^2 &  \rho\\
                         .& .& . & 1&    \rho\\
.& .& . & .&    1\\
 \end{pmatrix} \nn    \fourl \fourl 
\begin{pmatrix} \n 1\n & \n \rho_1 \rho_2 \n & \n  \rho_1 \rho_3 \n &\n   \rho_1 \rho_4 \n&  \rho_1\\
                         .& 1   &  \rho_2 \rho_3  &   \rho_2 \rho_4  &  \rho_2\\
                         .& .& 1 &  \rho_3 \rho_4 &  \rho_3\\
                         .& .& . & 1&    \rho_4\\
.& .& . & .&    1\\
 \end{pmatrix}.$
\end{table}

Another feature of the joint probabilities in \eqref{jointprobs} is that the conditional odds-ratios for each  pair $A_q, L$ given the remaining 
$Q-1$ variables  are equal to $\alpha^2$. When one interprets these as equal distances,  concentric rings such as those on the 
right of Figure 1 may result.  The number of rings increases with an increase of $Q$ as illustrated
with  Table~\ref{tab:intp}  in  Section 3.   This explains  the chosen name of this family of distributions. First, we generate the distributions  over  star graphs.

\section{Generating  a concentric-ring model over a star graph}

To shorten descriptions and notation, we call both, the outer nodes of the star graph   and the corresponding binary variables  $A_1, \ldots, A_Q$,  {\bf
leaves} and identify them sometimes by their indices $1, \dots, Q$. 
Similarly we call both the inner node of the star graph, and variable $L$, the {\bf
root}.
Often the root is unobserved, that is latent or hidden, and one main task is then to estimate the joint $p$-dimensional distribution from observations on only  the $Q$ leaves.

 The first main feature of a  joint distribution of concentric rings is mutual conditional independence
of the leaves given the root, written as

\begin{equation}   \label{cind}
(1  \ci 2 \, \ci \, \ldots \, \ci  Q) \, |  L. \end{equation}

Any density  generated over a star graph, irrespective of the types of variables,  is defined by $Q$ conditional densities, $f_{q|L}$, and a marginal density, $f_L$,  of the root. In the condensed node notation, with node set  $N=\{1, \ldots, Q, L\}$ of size $p$, 
the joint density
$f_N$  factorizes as

\begin{equation}
f_N = f_{1|L}\ldots f_{Q|L} f_L  \\.\label{factd}
\end{equation}

For binary variables, $f_N$ denotes the joint probability distribution, so that equation \eqref{factd} becomes 
$\pi(a_1,\ldots, a_Q,l)= \pi(a_1| l)\cdots \pi(a_Q| l)\pi(l) $
where $\pi(a_q| l)= \pi(a_q,  l)/\pi(l)$  are obtained from the bivariate probabilities  of each leave, $A_q$, and the root, $L$. In Table~\ref{tab:measures}, we show probabilities  for any binary pair $(A,L)$ and for each variable of the the pair being symmetric.

\begin{table}[H]
\caption{A $2\times 2$ table of a general  $(A,L)$ and in the special case of symmetric binary  variables \vspace{-2mm}}\label{tab:measures} 
\vspace{2mm}
\centering
\small
\begin{tabular}{l c cc c  l l c cc cc}
 \cline{1-5}  \cline{7-11}\\[-3mm]
&& \multicolumn{2}{c}{$L$} & & & & &\multicolumn{2}{c}{$L$}\\
\cline{3-4} \cline{9-10 }\\[-3mm]
$A$ && weak & strong&  \n sum  \n &\nn \nn  &$A$ && weak  & strong &sum\\
\cline{1-5}  \cline{7-11}\\[-3mm]
miss && $\pi_{\bm {mw}}$&  $\pi_{\bm{ms}}$  & $\pi_{\bm m}$&    & miss &&$\fracsfourth(1+\rho) $&$\fracsfourth(1-\rho) $   &\fracshalf\\[2mm]
succeed && $\pi_{\bm{sw}}$&  $\pi_{\bm ss}$&  $1-\pi_{\bm m}$ & & succeed && $\fracsfourth(1-\rho) $ & $\fracsfourth(1+\rho) $ & \fracshalf\\[2mm]
\cline{1-5}  \cline{7-11}\\[-3mm]
sum& &  $\pi_{\bm w}$ &  $\pi_{\bm s}$ & 1&&sum&& \fracshalf& \fracshalf &\n 1 \\[1mm]
 \cline{1-5}  \cline{7-11}\\[-3mm]
\end{tabular}
\end{table}

Several standard measures of dependence, that are in common use, are defined in Table~\ref{tab:depm} by using  Table \ref{tab:measures} and  equation \eqref{rhoalpha}, both for
a general binary  pair $(A,L)$ and for it being symmetric.

\begin{table}[H]
 \caption{Measures in a general $2\times 2$ table and  in the special case of symmetric binary  variables \vspace{-2mm}}\label{tab:depm} 
 \vspace{2mm}
\centering
\small
\begin{tabular}{lllllllllllllllllllll}
\toprule\\[-3mm]
definition &\multicolumn{17}{l}{interpretation in general and for  two symmetric binary variables\nn \nn \nn \n}\\[1mm]
\midrule\\[-4mm] 
$\pi_{\bm {ss}}/\pi_{\bm {ms}}$ & \multicolumn{11}{l}{odds of succeeding versus missing given a strong signal:\!\!}&$\alpha$\\
$\pi_{\bm {sw}}/\pi_{\bm {mw}}$ & \multicolumn{11}{l}{odds of succeeding versus missing given a weak signal:}&$1/\alpha$\\[1mm]
$(\pi_{\bm {ss}}\pi_{\bm {mw}})/(\pi_{\bm {ms}}\pi_{\bm {sw}})\!\!$& \multicolumn{11}{l}{odds-ratio for success or cross-product ratio:}&  $\alpha^2$\\[1mm]
$\pi_{\bm {s|s}}=\pi_{\bm {ss}}/  \pi_{\bm s}$&  \multicolumn{11}{l}{chance to succeed given a strong signal of $L$:}& $(1+\rho)/2$\\
$\pi_{\bm {s|w}}=\pi_{\bm {sw}}/  \pi_{\bm w}$& \multicolumn{11}{l}{chance to succeed given a weak signal of $L$:}& $(1-\rho)/2$\\[1mm]
$\pi_{\bm {s|s}} - \pi_{\bm {s|w}}$& \multicolumn{11}{l}{chance difference in succeeding:}  &$\rho$\\[1mm]
$\pi_{\bm {s|s}} /\pi_{\bm {s|w}}$& \multicolumn{11}{l}{relative chance for  success:} &$\alpha$ \\[1mm]
\bottomrule
\end{tabular}
\end{table}
 For both $A, L$ symmetric, 
the  parameter $\rho>0$   relates  also  directly to  the probabilities  via 
 \begin{equation} \rho=(\pi_{\bm{ss}}+\pi_{\bm{mw}})-(\pi_{\bm{sw}}+\pi_{\bm{ms}}),  \label{rhoallpi}\end{equation}
and  the odds of succeeding versus 
 missing given a strong signal of $L$  coincides with the relative chance for success. 
Independence of  any binary  pair $(A,L)$ requires in general, that the odds-ratio  equals one,  the relative chance equals one and the chance difference equals zero.

For the relation of $\alpha$ to conditional independence given $L$, we only look at pair $(A_1, A_2)$
at both levels of $L$ in Table~\ref{tab:pairgL}, since the mutual independence in equation \eqref{cind} implies independence of each pair of leaves from the remaining $Q-2$ leaves given $L$, in particular $(1,2)\ci (3,\dots,Q)|L$.

The conditional independence $1\ci 2|L$ is  directly reflected in the equal-one odds-ratios 
within the subtables for each level  of  $L$. The same holds for the relative chances, while  the chance difference and the correlation 
coefficient  in each subtable  for $(A_1,A_2)$ are zero  for $1\ci 2|L$. 
Table~\ref{tab:pairgL} shows in addition the {\bf
joint symmetry of the distribution}  since the probability for any  given level combination of the variables remains unchanged after switching all the levels. 
\begin{table}[H]
\caption{Probabilities with  $1\ci 2|L$ multiplied by $c_2=2(1+\alpha)^2$  for pair $(A_1,A_2)$ given $L$\vspace{-3mm}}\label{tab:pairgL} 
\vspace{2mm}
\centering
\small
\begin{tabular}{l c cc c ccc}
\toprule\\[-4mm]
&& \multicolumn{2}{c}{weak $L$} && \multicolumn{2}{c}{strong $L$}\\
\cline{3-4} \cline{6-7}\\[-3mm]
$A_1$ && $A_2$ miss & $A_2$ succeed & \nn \nn& $A_2$ miss& $A_2$ succeed &\nn sum\\
\midrule\\[-3mm]
miss && $\alpha^2 $&  $\alpha$&  &  1  & $\alpha^{\n} $ &$\nn \nn(1+\alpha)^2$\\ 
succeed && $\alpha^{\n}$&  1& & $\alpha$&  $\alpha^2 $&$\nn \nn (1+\alpha)^2$\\
\midrule\\[-3mm] 
sum &&$\alpha(1+\alpha)$ &  $(1+\alpha)$& & $(1+\alpha)$ &  $\alpha(1+\alpha)$& $\n \nn 2(1+\alpha)^2$\\
\midrule\\[-3mm]
odds-ratio&& \multicolumn{2}{c} {1}&& \multicolumn{2}{c} {1}\\
\bottomrule
\end{tabular}
\end{table}

Joint symmetry also holds in general, as can be derived directly from (\ref{jointprobs}). It  follows that the marginal distribution of each $(A_q,L)$ is symmetric and does not depend on $q$. For $\{-1,1\}$ coding, we have then from this symmetry and equation \eqref{factd}, for  the generated joint distribution in \eqref{jointprobs} and with $q=1, \dots, Q$
\begin{equation} \pi(a_1,\ldots,a_Q,l)\,\,\,=\,\,\,2^{-p}\,\txt{\prod}_{q} \,(1+ \rho \; a_{q}l).  \label{jointprobs1}
\end{equation}
 \section{Kronecker product representations of joint probabilities}
 We now  introduce 
for  $p\geq3$ a  vector representation.  For this, we write for instance 
$  \pi_{111}=\,pr
(A_1=1, A_2=1, L=1).$ 
 Then, by
 using again $N=\{1, \dots, Q, L\}$  and the $\{0, 1\}$ coding  
 and  letting the levels of the first variable change fastest, the column vector of probabilities, ${\bm \pi}
_{3, N}$,  is in transposed form 
\begin{eqnarray*}{\bm \pi}\T
_{3, N}&=&( \pi_{000},  \n \pi_{100},  \n \pi_{010}, \n  \pi_{110}, \n  \pi_{001},  \n \pi_{101},  \n \pi_{011}, \n  \pi_{111})\\ \n&=&
(\alpha^2 \!, \n \alpha, \n \alpha, \n1,  \n1, \n \alpha, \n \alpha, \n \alpha^2) /c_2, \end{eqnarray*}
where $c_2=2(1+\alpha)^2$ and we  take in this notation always the last variable  to coincide with $L$.

 For an integer-valued $\alpha$,   we illustrate next how the concentric rings are generated and increase
with the number of variables. One way  to generate the probabilities after an increase from  $p$ to $p+1$ nodes, is to
start with the probabilities at  the strong signal of  $L$ for the given $p$,
multiplied by $c_Q=2(1+\alpha)^{Q}$, to obtain first a vector of powers of $\alpha$ such as in Table~\ref{tab:intp}.

This vector is appended next by the same vector modified just by increasing the power of each $\alpha$  by one.
The joint probabilities for a strong signal of $L$ for $p+1$ nodes result after dividing by the new normalizing constant $c_{(Q+1)}=2(1+\alpha)^{Q}$
and repeating the probabilities in reverse order for the lower half  of the table.

\begin{table}[H]
\caption{Integer parametrization of the upper half of the probability vector  for $p=1$ up to $p=5$ variables;
with the sum of the integers equal to $2(1+\alpha)^{Q}$, the number of leaves equal to $Q=p-1$}\label{tab:intp} 
\vspace{2mm}
\centering
\begin{tabular}{clllllllllllllllllll}
\toprule
\\[-3mm]
$p$&  \multicolumn{16}{c}{moving from $p$ to $p+1$ using powers of $\alpha$}\\[1mm]
\midrule\\[-3mm]
1& $\alpha^0$  \\
2&  $\alpha^0$ & $\alpha^1$ \\
3&   $\alpha^0$& $\alpha^1$ & $\alpha^1$ & $\alpha^2$\\
4&   $\alpha^0$ & $\alpha^1$ & $\alpha^1$ & $\alpha^2$&  $\alpha^{1}$ & $\alpha^2$ & $\alpha^2$ & $\alpha^3$\\
5&   $\alpha^0$ & $\alpha^1$ & $\alpha^1$ & $\alpha^2$&  $\alpha^{1}$ & $\alpha^2$ & $\alpha^2$ & $\alpha^3$&
$\alpha^1$ & $\alpha^2$ & $\alpha^2$ & $\alpha^3$&  $\alpha^{2}$ & $\alpha^3$ & $\alpha^3$ & $\alpha^4$\\
\bottomrule
\end{tabular}
\end{table}

For large $p$, the row vector ${\bm \pi}_{p, N}\T
$ has a computationally attractive representation in terms of  Kronecker products. Let  $v=(1, \n  \alpha)$,  $w=( \alpha, \n 1)$ and $c_Q= 2(1+\alpha)^{Q}$, then
 ${\bm \pi}_{p, N}\T$ may be obtained from
 \begin{equation}
(\,\underbrace{w \otimes  \cdots \otimes w}_{p-1} \,  , \n \underbrace{v \otimes  \cdots \otimes v\,}_{p-1}  )/c_Q \label{kronp}\; .\end{equation}
 
 From the given form of the joint distribution, it can be checked directly that for any $p > 2$ and any  selected pair $(A_q,L)$, the conditional cross-product ratios equal  $\alpha^2$, the conditional relative chances for success equal $\alpha$ and the conditional chance differences in succeeding equal $\rho$, that is in all subtables formed by the level combinations of the remaining leaves.
 
 Collapsibility results for the three  measures show that these three measures remain unchanged after marginalizing over some or all of the remaining  leaves if these are conditionally independent of  $A_q$ given $L$; see Wermuth (1987) and  Xie, Ma and Geng (2008). The common strength of dependence of  each  $A_q$ on $L$ gives an increase of the number of  concentric rings as $p$ increases. 

 To compute moments  and other 
 features of the distribution in a fast way,  we show in the next section that Kronecker  products based on special $2\times 2$ matrices  are particularly helpful,
 since for instance the inverses of such products are the Kronecker products of the inverses.
 
 \section{Moments, interactions and sums of level  combinations of the leaves}
 The $\{0,1\}$ coding of binary variables is well suited to understand the change from raw and from central moments, in general,  to those of the concentric-ring model. With 
$$ \bcal_p=\underbrace{\bcal \otimes \cdots \otimes \bcal}_p, \nn \nn \bcal=\begin{pmatrix} 1 &\n 1\\ 0&\n 1\\\end{pmatrix},$$
the column vector of raw moments is, in general binary-star-graph models,
\begin{equation}  \bm{m}_{p,N} =\bcal_p \;{\bm \pi}_{p, N} \, . \label{rawmom}\end{equation}
For the concentric-ring distribution and $p=3$, the raw moments in  \{0 1\} coding reduce barely, with e.g. $  \pi_{11+}=\,pr
(A_1=1, A_2=1)=\txt{\sum}_{l}\pi_{11l}$ and  $ \pi_{1++}=\pi_{11+}+\pi_{10+}\, ,$ as follows, 
\begin{eqnarray*}
\bm{m}_{3,N}\T
&=&(1,  \n  \pi_{1++}, \n \pi_{+1+}, \n \pi_{11+},   \n  \pi_{++1}, \n \pi_{1+1}, \n \pi_{+11}, \n \pi_{111})\\
\n&=&(1,  \n  \fracshalf, \n \fracshalf, \n \beta,   \n  \fracshalf, \n \gamma, \gamma, \n \delta),
\end{eqnarray*} 
where $\beta=(1+\alpha^2)/c_2$, \, $\gamma=\alpha(1+\alpha)/c_2$, \, $\delta=\alpha^2/c_2$, \, $c_2=2(1+\alpha)^2$.

Another Kronecker product leads to central moments  of  $\{0,1\}$-coded binary variables; see Teugels (1990). For instance with $q=1, \ldots, Q$  and ${\bm \tcal\!}_{p,N}=\bm{\tcal\!}_1 \otimes \cdots \otimes  \bm{\tcal\!}_Q \otimes \bm{\tcal\!}_L$, where
$$    \bm{\tcal\!}_q =\! \begin{pmatrix} 1& 1\\-\,pr
(A_q=0 )& \,pr
(A_q=1) \end{pmatrix}  \nn  \bm{\tcal\!}_L =\!\begin{pmatrix} \nn 1& 1\\-\,pr
(L=0)& \,pr
(L=1) \end{pmatrix}\, , $$
the vector of central moments is
\begin{equation}  \bm{\mu}_{p,N} =\bm{\tcal\!}_{p,N} \;{\bm \pi}_{p, N} \, .\label{cenmom} \end{equation}
For concentric-ring distribution and  $p = 3$, the central moments reduce with $\gamma=\rho/4$ to 
\begin{eqnarray*}
\bm{\mu}_{3,N}\T
&=&(1,  \n  0, \n 0, \n \mu_{12},   \n  0, \n \mu_{13}, \n \mu_{23}, \n \mu_{123})\\
\n&=&(1,  \n  0, \n 0,\n  4\gamma^2,   \n  0, \n \gamma, \n \gamma , \n 0)  \, .
\end{eqnarray*} 
By the mixed-product property of  Kronecker products, simple relations result, such as for instance 
$$   \bm{\mu}_{p,N}=\bm{\tcal\!}_1 \bm{ \bcal}^{-1} \otimes \cdots \otimes \bm{\tcal\!}_Q \bm{ \bcal}^{-1}\otimes \bm{\tcal\!}_L \bm{\bcal}^{-1}  \, \bm{m}_{p,N}\, .\label{logint}
$$

By contrast, the  $\{-1,1\}$ coding of  binary variables is  well suited  to express the change from general log-linear interactions  to those that are much simpler in the concentric-ring model. With
$${\bm  \ecal}_p=\underbrace{{\bm \ecal} \otimes \cdots \otimes {\bm \ecal}}_p, \nn \nn {\bm \ecal}=\begin{pmatrix} 1 &\nn1\\ 1&-1\\\end{pmatrix},$$
the vector of  log-linear interactions for  the probabilities, at the combinations of levels  one, are
\begin{equation}  \bm{\lambda}_{p,N} =\underbrace{{\bm \ecal}^{-1} \otimes \cdots \otimes {\bm \ecal}^{-1}}_p\, \log({\bm \pi}_{p, N}) \label{loglinint} \,.\end{equation}
For the concentric-ring distribution and $p=4$, the log-linear interactions reduce   as follows
\begin{eqnarray*}
\bm{\lambda}_{4,N}\T
\!&=&\!(\lambda_{-}, \,
  \lambda_1,\,
 \lambda_2,\,
 \lambda_{12},  \,
  \lambda_3,\,
 \lambda_{13},\,
 \lambda_{23},\,
 \lambda_{123},\n  \lambda_4, \,
  \lambda_{14},\,
 \lambda_{24},\,
 \lambda_{124},  \,
  \lambda_{34},\,
 \lambda_{134},\,
 \lambda_{234},\,
 \lambda_{1234})\\
\n&=&\!(\beta,  \n  0, \n 0, \n 0,   \n  0, \n  0, \n  0, \n  0, \n  0, \n \gamma, \n \gamma, \n 0, \gamma,  \n 0, \n 0, \n 0),\end{eqnarray*} with $\gamma=\fracshalf\log(\alpha),\n \beta=3\gamma-\log(c_3)$.

In general, only the 2-factor terms that include $L$ and  the  overall normalizing constant $\lambda_{-}$ are nonzero.   In the log-linear parametrisation,
conditional independence of any pair  implies that all higher-order interaction terms involving this pair are vanishing as well; see e.g. Fienberg (2007). Thus,  the independences  of equation \eqref{cind} lead to all other log-linear interaction terms being zero.

For binary variables, the linear interactions may in general be defined with the same Kronecker product matrix as used for the log-linear interactions in equation \eqref{loglinint}
\begin{equation}  \bm{\xi}_{p,N} ={\bm \ecal}_p\; {\bm \pi}_{p, N} \, . \label{linint}\end{equation}
 These linear interactions reduce for the concentric-ring distribution and  $p=4$ as follows
\begin{eqnarray*}
\bm{\xi}_{4,N}\T
&=&(1,  \n  \xi_1, \n \xi_2, \n \xi_{12},   \n  \xi_3, \n \xi_{13}, \n \xi_{23}, \n \xi_{123}, \n \xi_4,  \n  \xi_{14}, \n \xi_{24}, \n \xi_{124},   \n  \xi_{34}, \n \xi_{134}, \n \xi_{234}, \n \xi_{1234})\\
\n&=&(1,  \n  0, \n 0, \n \rho^2 \!,   \n  0, \n \rho^2 \!, \n   \rho^2\!, \n  0, \n 0,  \n  \rho, \n \rho, \n 0, \n  \rho,  \n 0,\n  0, \n \rho^3)\, .
\end{eqnarray*}

From equations \eqref{cenmom},
and \eqref{linint} and from $\bm{\ecal \tcal}_q^{-1}$ being of diagonal form,  the linear-interaction terms in $\bm{\xi}_{p,N}$ are just rescaled versions of 
the central moments  $\bm{\mu}_{p,N}$. They are the standardized central moments, that result after transforming the $\{0,1\}$ coded binary variables $A_q$, say, with mean $\fracshalf$ and variance $\fracsfourth$, into their standardized form with $\{-1,1\}$-coding, that is into  $A^*_q =2A_q-1$.

This central moment  representation is more complex than the log-linear formulation because of the non-vanishing 4-factor interaction term $\rho^3$. For $Q>3$, each odd-order interaction is zero, an even-order $k$-factor interaction involving the root as the last variable, is $\rho^{k-1}$ and it is $\rho^{k}$, otherwise. As we shall see, this advantage of the log-linear interactions disappears in the marginal distribution of the leaves which has no independences.  

For later use, we introduce  the sum $S$, and the average $\bar{S}$, of the $Q$ standardized
variables  $A_q^*$. Under the concentric ring model, each
pair has the same correlation $\rho^2$, see Table  \ref{tab:cormat}, so that
\begin{equation} {\var(S)=Q+ 2\small{{Q \choose 2}}\;\rho^2,    \nn \nn Q\,\var(\bar{S})=1+ (Q-1)\;\rho^2.
 \label{varS}}
 \end{equation}

Also directly from the right of Table \ref{tab:measures}, one sees that
$E(A_q^* | L = 0) = -\rho$ and $ E(A_q^* | L = 1) = \rho$ so that
\begin{equation}
E(\bar S|L = 1) - E(\bar S|L = 0) = 2  \rho.  \label{diffSgL}
\end{equation}
\section{Marginal distributions of the leaves}
By the joint symmetry, marginalizing over the common root returns a symmetric distribution by construction.
This is illustrated in  Table~\ref{tab:margleaves} for $Q=3$ leaves.\vspace{-3mm}
\begin{table}[H]
\caption{Marginalising over $L$  for $Q=3$ by adding $\alpha$'s for corresponding  level combinations of the leaves;
in the table each probability is multiplied by $c_3=2(1+\alpha)^3$}\label{tab:margleaves} 
\vspace{1mm}
\centering
\begin{tabular}{l cc cc cc cc cc }
\toprule\\[-4mm]
$2^3$ levels: &000& 100& 010& 110& 001& 101& 011& 111\\
\midrule\\[-4mm]
at  level $l=0:$&$\alpha^0$ & $\alpha^1$ & $\alpha^1$ & $\alpha^2$&  $\alpha^{1}$ & $\alpha^2$ & $\alpha^2$ & $\alpha^3$\\
at level $l=1:$&$\alpha^3$ & $\alpha^2$ & $\alpha^2$ & $\alpha^1$&  $\alpha^{2}$ & $\alpha^1$ & $\alpha^1$ & $\alpha^0$\\ \midrule\\[-4mm]
margin over $L: \!\!\!\!$&$1+\alpha^3 \!\!$&$\alpha+\alpha^2 \!\!$& $\alpha+\alpha^2 \!\!$&$\alpha+\alpha^2 \!\!$&$\alpha+\alpha^2 \!\!$&$\alpha+\alpha^2 \!\!$& $\alpha+\alpha^2 \!\!$&$1+\alpha^3 \!\!$\\
\bottomrule
\end{tabular}
\end{table}

In general, after marginalizing over $L$,  the distribution of the remaining $Q$ leaves  is given, with $K_t$ denoting again the number of ones in
any  sequence of levels, $(a_1,\dots, a_Q)$, by
\begin{equation}\pi(a_1,\dots ,a_Q)=\frac{1}{c_Q}\left(\alpha^{K_t}+\alpha^{(Q-K_t)}\right). \label{margprobs}
\end{equation}
One also obtains the linear interaction vector for the joint distribution of the leaves,
$\bm{\xi}_{p,N\setminus L}$  as the lower half of  $\bm{\xi}_{p,N}$,
where  in this notation, we do not distinguish between an element $L$ and  the singleton $\{L\}$.  

Equivalently, with  $\xi_{0}=1$ and $\ical\subseteq\{1,\ldots, Q\}$ such that $q \in \ical$ if and only if $a_q=1$ is  in  $(a_1, \ldots, a_Q)$, as used before in the example to equation \eqref{linint}, the other elements of $\bm{\xi}_{p,N\setminus L}$ may be written as
\begin{equation} \label{xim}
\xi_{\ical}=
\begin{cases}
\rho^{K_t} &  \text{for even } K_t ,\\
0 & \text{otherwise\, .} \\
 \end{cases}
\end{equation}
Also  
$ \bm{\lambda}_{p,N\setminus L}=\bm{\ecal}^{-1}_{Q}
\log\{(\bm{\ecal}_{Q})^{-1} \bm{\xi}_{p,N\setminus L} \}        $ has zero values  in the same positions as   $\bm{\xi}_{p,N\setminus L} $.
Thus, all odd-order log-linear interactions vanish and all odd-order (standardized) central moments vanish. The even-factor terms are
functions of $\rho^2$ which is the induced marginal correlation for any pair of leaves  and, at the same time, the  induced difference in chances for 
success.

\section{The conditional distribution of the root given the leaves}

From equations \eqref{jointprobs} and \eqref{margprobs},  the conditional distribution, $\pi(l | a_1,\dots ,a_Q)$, of  the root, $L$, given the leaves, $A_1, \dots A_Q$, satisfies
 in  terms of $\alpha$,  the number of ones, $K_t$,  in  the leaf-level sequence $(a_1, \dots, a_Q)$\begin{equation} \label{condLprobs}
c_Q\,  \pi(a_1,\dots ,a_Q)\, \pi(l | a_1,\dots ,a_Q)   =
\begin{cases}
\alpha^{K_t}&  \text{ for } l=1,\\
\alpha^{(Q-K_t)}&  \text{ for } l=0 \, .\\
 \end{cases} 
\end{equation}

Functions of the odds-ratio are known to be the only measures of dependence in $2\times 2$ tables that are variation independent
of the margins; see Edwards (1963). By using the concentric-ring model, we  illustrate now  how  the relative chance 
and the chance difference may give strongly distorted impressions of equal conditional dependences. 

When the roles of   explanatory variable $L$ and responses  in the given generating process are exchanged,
 the odds-ratios stay constant,
equal chance differences  appear to be of sharply reduced strengths and equal relative risks appear to be  strongly unequal.

To see this, we compare the dependences of $A_2$ on
$L$ given $A_1$ using the odds-ratio, odr$(A_2, L| A_1$), the chance differences  to succeed, chd$(A_2, L| A_1$), and the
relative chances to succeed, rch$(A_2, L| A_1)$, with the  corresponding dependences of  $L$ on $A_2$  given $A_1$.
After  exchanging the ordering to ($A_2$, $L$, $A_1$) in  Table~\ref{tab:pairgL} and taking as an example $\alpha=9$: 
$$  \text{odr}(A_2, L\,| A_1=a_1)=\alpha^2=81 , \n \text{chd}(A_2, L| A_1=a_1)=\rho=0.80, \n \text{rch}(A_2, L| A_1=a_1)=9,$$
while from Table~\ref{tab:LA_2gA_1} with $L$ as the first variable and $A_2$ as the second, one obtains
$$  \text{odr}(L, A_2\,| A_1=1)=81 , \fourl  \n \text{chd}(L, A_2\,| A_1=1)=0.49, \fourl \text{rch}(L, A_2\,| A_1=1)=41,$$
$$\nn   \text{odr}(L, A_2\,| A_1=0)=81 , \fourl \n \text{chd}(L, A_2\,| A_1=0)=0.49, \fourl \text{rch}(L, A_2\,| A_1=0)=1.98.$$

 \begin{table}[H]
\caption{Probabilities multiplied by $2(1+a)^2$ for $(A_1,A_2,L)$ in Table~\ref{tab:pairgL} reordered as $(L, A_2, A_1)$}\label{tab:LA_2gA_1} 
\vspace{2mm}
\centering
\small
\begin{tabular}{l c cc c ccc}
\toprule\\[-4mm]
& \multicolumn{2}{c}{$A_1$ miss } && \multicolumn{2}{c}{$A_1$ succ.}\\ 
\cline{2-3} \cline{5-6}\\[-3mm]
level $l$ of $L$ & $A_2$ miss & $A_2$ succeed & \n & $A_2$ miss& $A_2$ succ. \\\\[-3mm]
\midrule\\[-3mm]
0 := weak & $\alpha^2 $&  $\alpha^{\n}$&  &  $\alpha^{\n}$  & 1  \\ 
1 := strong & 1 &  $\alpha^{\n}$& & $\alpha^{\n}$&  $\alpha^2 $\\
\midrule\\[-3mm] 
sum &$(1+\alpha^2)$ &  $2\alpha \nn $& & $2\alpha \nn$ &  $(1+\alpha^2)$\\
\midrule\\[-3mm]
odds-ratio for $l=1$, $a_2=1$; odr$(L,A_2|A_1)$& \multicolumn{2}{c} {$\alpha^2$}&& \multicolumn{2}{c} {$\alpha^2$}\\[1mm]
relative chance for $l=1$; rch$(L,A_2|A_1)$& \multicolumn{2}{c}{$(1+\alpha^2)/2$}&& \multicolumn{2}{c}{$2\alpha^2/(1+ \alpha^2)$}\\[1mm]
chance difference  for $l=1$; chd$(L,A_2|A_1)$\!\!\!& \multicolumn{2}{c}{$\fracshalf-1/(1+\alpha^2)$}&& \multicolumn{2}{c}{$\alpha^2/(1+\alpha^2)-\fracshalf$}\\
\bottomrule
\end{tabular}
\end{table}

We notice next that in a logit regression of $L$ on the leaves, $A_q$, the regression parameters are functions of the conditional odds-ratios for $(L, A_q)$ since  they may be  obtained from  twice the log-linear 
parameters $\bm{\lambda}_{p,N}$ {in equation \eqref{logint} that do not involve $L$. This follows from 
the definition of the joint probabilities in  \eqref{jointprobs} and  the logit representation 
$$ \logit\{\pi(l | a_1,\ldots a_Q)\}= \log \pi(a_1,\ldots, a_Q, 1) - \log  \pi(a_1,\ldots, a_Q, 0).
$$

Thus, the odds-ratio and this logistic regression coefficient  are unaffected by switching the roles of $A_2$ and $L$, while the strength
of dependence measured with the chance difference is reduced in the example from 0.80 by almost 40\%  to 0.49 and the dependences 
measured with equal  relative chances of 9 for $A_2$ on $L$ , are modified into 41 and about 2, thus clearly into  strongly different strengths 
of dependence at the two levels of $A_1$. 
As $Q$ increases, the relative chance for a strong signal, comparing succeeding to missing
in $A_1$, increases even to $(1+\alpha^Q)/2$ at $Q-1$ misses of the remaining variables.

 Such  changes illustrate  potential problems for machine learning and causal conclusions,  for  interpretations of  some 
case-control studies and for some uses of the propensity score.

\section{Maximum-likelihood estimates}
One of the most attractive properties of the maximum-likelihood estimate  of a set of parameters in a given model is that the maximum-likelihood estimate  of any other set of parameters, related to the original ones by a one-to-one (1--1) transformation, is given by the same 1--1 transformation for the estimates; see Fisher (1922).
Thus here, given the maximum-likelihood estimate $\hat \rho$ of $\rho$,  all other measures of dependence are defined by the relevant 1--1 transformations. Given $\hat{\alpha}$, the maximum-likelihood estimates of the log-linear interactions are also given. Furthermore,  other estimated interactions of interest,  as well as the joint probabilities, result via the 1--1 transformations of Section 4.

Given the observed frequencies, for a  pair $(A,L) $ of  symmetric binary variables  that sum to $n$
in vector $ {\bm n}\T
_{2, N}=(n_{00}, \n n_{10}, \n n_{01},\n n_{11})$,  one obtains with equations \eqref{jointprobs1} and  \eqref{rhoallpi}
 \begin{equation}
\hat \rho= \{(n_{00} + n_{11} )- (n_{01} + n_{10}  )  \}/n := \text{ csd}_{AL}, \label{csd}
\end{equation}
where `csd' abbreviates  `{\bf
cross-sum difference}', a term introduced by G.M. Marchetti  in recent  unpublished work. 
For  symmetric variables $A_1, A_2, L$   observed and  satisfying $1 \ci 2 \mid L$ and $\E(A_1\,L)=\E(A_2\, l)=\rho$,  given the vector of counts $ {\bm n}\T
_{3, N}$,
we get  the average of the two cross-sum differences as the  unique maximum-likelihood estimate
$$  
\hat \rho=\fracshalf(  \text{csd}_{1L} +  \text{ csd}_{2L})
$$
of the common correlation.
Similarly, for  observations $n(a_1, \dots, a_Q, l)$ on $A_1,\dots, A_Q, L$   of a concentric-ring model, the closed-form maximum-likelihood estimate  of $\rho$ equals the 
average of the $q=1, \dots, Q$ cross-sum differences in counts for each leaf-root pair $(A_q, L)$:
\begin{equation}\hat{\rho}=\frac{1}{Q}\txt{\sum}_{q} \text{\,csd}_{qL}\,. \label{hatrho}
\end{equation}

When $L$ is hidden, it can be shown for $Q=2$, that the maximum-likelihood estimate of $\rho^2$ equals  the observed cross-sum difference, and  for 
$Q=3$, that  it equals the average of the  three observed  cross-sum differences. For $Q>3$, there is in general no closed-form solution 
of the likelihood equation to estimate $\rho^2$, but 
a method-of-moment estimator $\tilde{\rho}^2$ of $\rho^2$ is obtained from equation~\eqref{varS} as 
\begin{equation}
\tilde \rho^2  = (Q\, \hat v - 1)/(Q-1) \label{estrhosq}
\end{equation}
where $\hat v$ is any sample estimate of  $\var(\bar S)$, and $v$ and $\bar{S}$ are as defined for equation~\eqref{varS}.

An EM algorithm (Dempster, Laird and  Rubin, 1977) for $\rho$ in the concentric ring model can be defined with closed-form solutions both for the E(expectation) and
for  the M(maximization) steps. 
In an E-step,  the $2^p$ joint estimated counts $\tilde n(a_1, \dots, a_Q, l) $  are  from the observed  $2^Q$ marginal counts  
 of the leaves, $n(a_1, \dots, a_Q)$, and  the conditional distribution of the root given the leaves:
 $$
 \tilde n(a_1, \dots, a_Q, l) = n(a_1, \dots, a_Q) \tilde \pi(l|a_1, \dots, a_Q)
 $$
 using a current estimate of $\alpha$ and equation \eqref{condLprobs}. 
 In an M-step, the estimated correlation coefficient results with the new  $2^p$ joint counts $ \tilde n(a_1, \dots, a_Q, l)$ via equation \eqref{diffSgL}. 

Also, the two steps can be combined into a single updating equation for the correlation coefficient. 
For this, we denote by  $\rho(m)$ the value of the correlation coefficient at   iteration step $m$ and  start with an initial estimate from
equation \eqref{estrhosq},
$\rho(0) = (\tilde \rho^2)^{1/2}$. Let $n_t = n(a_1, \dots, a_Q)$ and 
$s_t = a^*_1 + \cdots + a^*_Q$ be the marginal counts and the associated 
sum in $\{-1,1\}$ coding, respectively. Then, the updated estimate $\rho(m+1)$  is, with $t=1, \dots, 2^Q$, such that\begin{equation}
\,\rho(m+1)\,\,=\frac{1}{n Q}\, \txt{\sum}_{t}  \,T_t(m)\,  n_t, \label{mstep}
\end{equation}
where we use  the relation between $\alpha$ and $\rho$ in equation \eqref{rhoalpha} to lead to 
\begin{equation}
T_t(m) = s_t
\{\alpha(m)^{\displaystyle{s_t}}-1\} /\{\alpha(m)^{\displaystyle{s_t}}+1\}.
\label{mstep2}
\end{equation}
Notice that, from a  table of counts for all $p$ variables and equation \eqref{diffSgL},
an estimate of  $\rho$ is  
$$
\large{\frac{1}{nQ}}\textstyle{\sum}_t \,s_t \,\{n(a_1, \dots, a_Q, 1) - n(a_1, \dots, a_Q, 0) \}.
$$
Then, from equations \eqref{margprobs} and \eqref{condLprobs}, at a given iteration of the EM algorithm, we can write 
$$
\tilde n(a_1, \dots, a_Q,1) - \tilde n(a_1, \dots, a_Q, 0)
 =\frac{\alpha^{\displaystyle{K_t}}- \alpha^{\displaystyle{(Q-K_t)}}}{\alpha^{\displaystyle{K_t}} +  \alpha^{\displaystyle{(Q-K_t})}} \,n_t   
 = \frac{\alpha^{\displaystyle{s_t}}-1}{\alpha^{\displaystyle{s_t}} + 1}\, n_t\,,
$$
so that equations \eqref{mstep} and \eqref{mstep2} follow.

To see that $T_t(m)$ in  equation \eqref{mstep} is for $\rho(0)>0$ always nonnegative, note that
 if $s_t \geq 0$ in equation \eqref{mstep2} then also
$(\alpha(m)^{\displaystyle s_t}-1)/(\alpha(m)^{\displaystyle s_t}+1)\geq 0$ because $\alpha(m)\geq 1$.
Similarly, if $s_t<0$ then $(\alpha(m)^{\displaystyle s_t}-1)/(\alpha(m)^{\displaystyle s_t}+1)\leq 0$.

The algorithm converges to a stationary point of the likelihood and 
the standard error of the estimate can be found using one 
of the methods discussed in \citet[Sect.~4.4, p.~74]
{Tanner96}. In extensive simulations under the model with 
$Q = 4$, the number of iterations required for 
convergence, for $\rho$ in the range of most interest, in
$0.5 < \rho < 0.8$,  was with a tolerance of $\epsilon = 10^{-4}$ at most 4 and 
with a tolerance  of $\epsilon = 10^{-7}$ at most 20. The absolute difference 
between $\hat{\rho}$ and $\rho$ was less 
than $0.1$ and less than $0.05$,  
in  samples of  size 300 and 1000 respectively.

\section{Discussion} 

A family of  jointly symmetric distributions in equally probable binary variables has been defined, where for  each given number of variables,
a distribution is characterized by a single parameter.  The family is shown to have several  attractive features that were not previously identified even though it is a special case of a number of models that have been intensively studied, such as Ising models of ferromagnetism, latent class structures and models for constructing phylogenetic trees. 

In particular, such a distribution is a graphical Markov model, generated over a star graph with $p-1$ leaves and one common root. A positive dependence 
of each leaf on  the root equals   a positive Pearson's correlation coefficient,  $\rho$.
When $p$ increases with $\rho$ kept fixed, the model leads to  an increasing number of evenly-spaced concentric rings.  

 An integer parametrization shows
which sample size is needed so that the smallest  count is expected to equal one. This information helps to plan for observed positive distributions, that is for a  
sufficient condition that the intersection property  (see e.g.\ Pearl, 1988) holds for a given set of  observations on symmetric binary variables.

A closed-form maximum-likelihood estimate  $\hat{\rho}$ of $\rho$ is  obtained when the root is observed in addition to the leaves. 
Otherwise, a closed form method-of-moment estimate  $\tilde{\rho}$ of $\rho$ is derived. This estimate is a good starting value for the EM algorithm which reduces  to a single updating equation to obtain  $\hat{\rho}$. Simulations suggest that  $\tilde{\rho}$ and $\hat{\rho}$ agree often up to the second decimal place, that
 the likelihood function for $\rho$ has a unique maximum and that
it is quite flat only for $\rho \leq 1/3$ that is for the rather small dependences among  each leave pair of  only $\rho^2\leq  1/9$. With $\hat{\rho}$ estimated just from observations 
on the leaves, the joint probabilities or interactions including the root are  available in terms of Kronecker products of small matrices even for many variables.

The models are also used to illustrate how conditional relative chances and  chance differences  can change strongly,
when the roles of a regressor variable and the response  are exchanged,  while 
odds-ratios and  logit regression coefficients  capture the unchanged equal dependences given the remaining leaves. This problem occurs more generally but is convincingly demonstrated using this special binary family of distributions.

For two  binary variables,  in general, Pearson's correlation coefficient, $\rho$, is a multiple of  the cross-product difference of the probabilities; see for instance equation (10)
in Wermuth and Marchetti (2014). Only for symmetric binary variables, $\rho>0$ reduces to the cross-sum difference in equation \eqref{rhoallpi}
and becomes a 1-1 function of the odds-ratio.
The cross-sum difference of counts in equation \eqref{csd}, arises  also as the nonparametric measure of dependence, studied by Blomqvist (1950) for continuous random variables:  in the special case  of symmetry in the observed $2\times 2$  table that may result after median-dichotomizing 
the bivariate observations. Extensions of this measure and relations to copulas have been investigated by Schmid and Schmidt (2007) and Genest, Carabarin-Aguirre and Harvey (2013).

The  one-parameter model considered here may be
generalized in several ways. One possibility is to abandon the
assumption of symmetry. For binary variables, this leads to the model studied for example in  Allman et al.\ (2014).
However even for 
this minimally extended model, it is much more complex to provide 
detailed insight  into  maximum-likelihood inference. 
In future work, we intend to study symmetric variables with more than two 
levels, concentric rings of binary variables with unequal spacings and maximization of  the empirical likelihood functions.


\section*{References}

\begin{thebibliography}{00}  \small
\bibitem[Allman et al.(2014)]{AllmanEtal14}
\textsc{Allman, E.S., Rhodes, J.A., Sturmfels, B. and Zwiernik, P.} (2014).
Tensors of nonnegative rank two.  ArXiv:1305.0539 and  \textit{ Linear Algebra and Applic.: Special Issue on Statistics.}
To appear.

\bibitem[Besag(1974)]{Besag74} 
\textsc{Besag, J.} (1974).
{Spatial interaction and the statistical analysis of lattice systems.}
\textit{J. Roy. Statist. Soc.  B}
\textbf{36}, 192--236.
  

\bibitem[Blomqvist(1950)]{Blomq50}
 \textsc{Blomqvist, N.} (1950). On a measure of dependence between two random variables. 
 \textit{Ann. Math. Stat.} \textbf{21},
593--600.


\bibitem[Castelo and Siebes(2003)]{CasSie03}
\textsc{Castelo, R. and  Siebes. A.} (2003).
A characterization of moral transitive acyclic directed graph Markov models as labeled trees.
\textit{J. Stat. Plan. Inf.}
\textbf{115}, 235--259.




\bibitem[Cox and Wermuth(1994)]{CoxWer94}
\textsc{Cox, D.R. and Wermuth, N.} (1994).
A note on the quadratic exponential binary distribution.
\textit{Biometrika} \textbf{81}, 403--406.
 
 \bibitem[Darroch, Lauritzen and Speed(1980)]{DarLauSpe80}
\textsc {Darroch, J.N., Lauritzen, S.L. and Speed, T.P.} (1980)  Markov fields and log-linear models for contingency tables. 
\textit{Ann. Statist.} {\bf 8}, 522--539.
  
  
 \bibitem[Dempster, Laird and Rubin(1977)] {DemLaiRub77} 
 \textsc{Dempster, A.P. , Laird, N.M., Rubin, D.B.} (1977). 
  Maximum likelihood from incomplete data via the EM algorithm.
  \textit{J. Roy. Statist. Soc. B}
  \textbf{39}, 1--38.
 
\bibitem[Edwards(1963)] {Edw63}
\textsc{Edwards, A. W. F.} (1963).
The measure of association in a 2 $\times$ 2 table.
\textit{J. Roy. Statist.  Soc.  A}
\textbf{126},
109--114.



 
\bibitem[Fienberg(2007)] {Fienb07}
\textsc{Fienberg, S.E.} (2007).
\textit{The Analysis of Cross-classified Categorical Data}, 2nd ed.
Springer, New York.
 
 \bibitem[Fisher(1922)]{Fisher22}
 \textsc {Fisher, R.A.} (1922).
  On the Mathematical Foundations of Theoretical Statistics.
  \textit{Philos. Trans. Roy. Soc. London Ser. A}
  \textbf{222}, 309--368.
   
   
   \bibitem[Genest, Carabarín-Aguirre, and Harvey]{Genestetal13}
   \textsc{Genest, C., Carabarín-Aguirre, A. and Harvey, F.} (2013).
   Copula parameter estimation
using Blomqvist's beta.
   \textit{J. Soc. Fran\c{c}. Statist.} 
\textbf{154}, 5--24.
   

 
\bibitem[Lazarsfeld(1950)] {Lazar50}
\textsc{Lazarsfeld, P.F.} (1950). The logical and mathematical foundation of latent structure analysis.
\textit{Measurement Prediction} \textbf{4}, 362--412.

\bibitem[Linzer and Lewis(2011)]{LinzLew11}
\textsc{Linzer, D.A. and Lewis J.B.} (2011).
{poLCA: An R package for polytomous variable latent class analysis}
\textit{J. Statist. Softw.}  \textbf{42}.
 
\bibitem[Pearl(1988)]{Pea88}
\textsc{Pearl, J.} (1988).
\textit{Probabilistic Reasoning in Intelligent Systems.}
Morgan Kaufmann, San Mateo.

   
\bibitem[Perlman and  Wu(1999)]{PerlWu}   
\textsc{Perlman, M. and Wu, L} (1999).    
 {Lattice conditional independence models for contingency tables with non-monotone missing data patterns.}
  \textit{J. Statist. Plan. Inference}
  \textbf{79}, 259--287.


\bibitem[Schmid and Schmidt(2007)]{SchmSchm07}
\textsc{Schmid, F.  and  Schmidt, R.} (2007).
Nonparametric inference on multivariate versions
of Blomqvist's beta and related measures
of tail dependence. 
\textit{Metrika} \textbf{66}, 323--354.

\bibitem[Tanner(1996)]{Tanner96}
  \textsc{Tanner, M.A.} (1996). \textit{Tools for Statistical Inference}, 3rd ed.
Springer, New York.

 \bibitem[Teugels(1990)]{Teug90}
 \textsc{Teugels, J.L.} (1990).
 Some representations of the multivariate Bernoulli and binomial distributions. 
 \textit{J. Multiv. Analysis}
 \textbf{32},  256-268.
    
\bibitem[Wermuth(1987)]{Wer87}
\textsc{Wermuth, N.} (1987). Parametric collapsibility and the lack of
moderating effects in contingency tables with a dichotomous response variable.
\textit{J. Roy. Statist. Soc. B}, {\bf 49}, 353--364.
 
\bibitem[Wermuth and  Lauritzen(1983)]{WerLau63}
\textsc{Wermuth, N. and Lauritzen, S.L.} (1983).
Graphical and recursive
models for contingency tab\-les. 
\textit{Biometrika}, {\bf 70}, 537--552.

 \bibitem[Wermuth and Marchetti(2014)]{WerMar14}
\textsc{Wermuth, N. and   Marchetti, G.M.} (2014).
Star graphs induce tetrad correlations:
for Gaussian as well as for binary
variables.
\textit{Electr. J. Statist.} \textbf{8}, 253--273.

 \bibitem[Wermuth, Marchetti and Cox(2009)]{WerMarCox09}
\textsc{Wermuth, N.,  Marchetti, G.M.  and Cox, D.R.} (2009).
Triangular systems for symmetric
binary variables.
\textit{Electr. J. Statist.}
\textbf{3}, 932--955.

\bibitem[Xie,Ma and Geng(2008)]{XieMaGeng08}
\textsc{Xie, X.C., Ma, Z.M. and Geng, Z.}  (2008). 
Some association measures and their collapsibility. 
\textit{Statist. Sinica}.  \textbf{18}, 1165--1183. 

 \bibitem[Zwiernik and Smith(2011)]{ZwierSmith11}
\textsc{Zwiernik, P.  and Smith, J.Q.} (2011).
 Implicit inequality constraints in a
binary tree model.
 \textit{Electr. J. Statist.}
\textbf{5}, 1276--1312.
\end{thebibliography}
\end{document}